\documentclass[letterpaper]{article} 
\usepackage{aaai2026}  
\usepackage{times}  
\usepackage{helvet}  
\usepackage{courier}  
\usepackage[hyphens]{url}  
\usepackage{graphicx} 
\usepackage{natbib}  
\usepackage{caption} 
\usepackage{algorithm}
\usepackage{algorithmic}
\usepackage{booktabs}
\usepackage{amsmath}
\usepackage{amssymb}
\usepackage{framed}

\newenvironment{promptbox}[1]{%
  \begin{framed}
  \noindent\textbf{#1}

  \smallskip
  \small
}{%
  \end{framed}
}

\usepackage{newfloat}
\usepackage{listings}
\DeclareCaptionStyle{ruled}{labelfont=normalfont,labelsep=colon,strut=off} 
\floatstyle{ruled}
\newfloat{listing}{tb}{lst}{}
\floatname{listing}{Listing}

\title{The Deliberative Deficit:\\
An Empirical Critique of LLMs in Democratic Discourse}
\author {
    Maurice Flechtner
}

\affiliations{
    University of Zurich, Zurich, Switzerland\\
    ETH Zurich, Zurich, Switzerland\\
    maurice.flechtner@zda.uzh.ch
}

\begin{document}

\maketitle

\begin{abstract}
Large language models (LLMs) are increasingly deployed in settings that require collective reasoning on complex, value-laden problems. Confidence in these deployments rests largely on benchmarks for \emph{verifiable} tasks (mathematics, coding, coordination games), yet many of these applications concern problems where no objectively correct answer exists and where decision quality instead depends on integrating pluralistic perspectives to find mutually acceptable solutions. We argue that LLM reasoning capacity on this class of problems cannot be fully inferred from verifiable-task benchmarks, and that procedural evaluations of LLM discourse (respectfulness, justification, engagement) are systematically insufficient. We apply the Deliberative Reason Index (DRI), a measure developed in political science and validated across citizen assemblies, as a tool for evaluating \emph{reliable group-level reasoning on pluralistic, non-verifiable problems}. Synthesizing recent evidence across 1{,}980 five-agent LLM runs on 12 citizen-assembly topics across 11 frontier model configurations, we find that LLM groups produce discourse with procedural quality comparable to human deliberation, while gains in intersubjective consistency are small, topic-dependent, and concentrated on tractable rather than ethically contested questions.
LLM groups exhibit roughly one-third the perspective diversity of human assemblies and reverse the human convergence pattern: human deliberation decreases dispersion as diverse views synthesise, whereas LLM deliberation increases it. Engineering diversity through persona prompting does not restore the human dynamic but inverts which component of deliberative reasoning is updated.
We argue this warrants a critique of the deliberative capacity of LLMs that is missing from current benchmarks and dangerous for deployment. Our conclusion is constraining rather than prohibitive: LLMs can function as tools supporting human reasoning on pluralistic problems, but current evidence does not license treating them as autonomous deliberative agents.
\end{abstract}

\begin{links}
    \link{Code (benchmark study)}{https://doi.org/10.1145/3772363.3798499}
\end{links}

\section{Introduction}

Large language models (LLM) are increasingly proposed for collective reasoning on consequential, value-laden problems. Recent proposals include AI-mediated consensus-finding in democratic deliberation~\cite{tesslerAICanHelp2024}, AI representation of missing perspectives in policy consultations~\cite{fulayEmptyChairUsing2025,zhuCanAITruly2025}, systematically AI-augmented citizen assemblies~\cite{landemoreCanArtificialIntelligence2024,mckinneyIntegratingArtificialIntelligence2024}, large-scale moderation of online deliberation~\cite{kleinModeratingLargeScale2025}, and simulations of democratic systems~\cite{rountreeCaseUsingGenerative2026, novelliReplicaOurDemocracies2025}. Proponents argue these systems could help scale the deliberative quality of small group deliberation to mass publics~\cite{landemoreCanArtificialIntelligence2024, lazarUsingLLMsEnhance2026} while sceptics worry about technosolutionism and the erosion of democratic substance~\cite{oleartWhyAITechnosolutionism2025, summerfieldImpactAdvancedAI2025}. However, as of now, the debate is missing a discussion of the question of whether LLMs can, in some meaningful sense, \emph{reason} about the kinds of problems deliberation addresses.

This premise is rarely tested directly. Confidence in LLM reasoning capacity rests primarily on benchmarks for tasks with verifiable solutions: mathematics and chess~\cite{duImprovingFactualityReasoning2023}, logical puzzles~\cite{liangEncouragingDivergentThinking2024, wuCanLLMAgents2025}, and coordination games~\cite{anneHarnessingLanguageCoordination2025}. Reinforcement learning with verifiable rewards (RLVR) has driven most recent progress in LLM reasoning~\cite{deepseek-aiDeepSeekR1IncentivizingReasoning2025}, and the field itself acknowledges that its advancements are mostly confined to domains with well-defined outcomes~\cite{zhangExtendingRLVROpenEnded2026a}.

Yet the collective reasoning problems for which LLMs are being deployed in democratic contexts are not verifiable tasks. Citizen assemblies on climate policy, stakeholder negotiations on resource allocation, participatory design processes, or ethics consultations on new technologies are all problems where no objectively correct answer exists, where legitimate perspectives diverge, and where decision quality depends on how well diverse considerations are integrated into a shared reasoning framework. Deliberation theory has long recognized that the goal of reasoning on such problems is not correctness but \emph{mutually acceptable, well-reasoned conclusions} that integrate pluralistic considerations~\cite{niemeyerEndsDeliberationMetaconsensus2007,dryzekReconcilingPluralismConsensus2006} and that the epistemic value of this integration depends constitutively on the diversity of perspectives brought into it~\cite{landemoreDeliberationCognitiveDiversity2013}. We refer to this class of non-verifiable, pluralistic and often value-laden problems as \emph{pluralistic reasoning problems}, and argue they constitute a significant challenge for LLM evaluation today.

When verifiable metrics do not apply, evaluation often falls back on \emph{procedural} measures. In the case of democratic deliberation, this would revolve around whether the system produces justified, respectful, reciprocal discourse. Such measures, as seen in the Discourse Quality Index~\cite{steenbergenMeasuringPoliticalDeliberation2003} and its automated variant AQuA~\cite{behrendtAQuACombiningExperts2024}, capture \emph{how} agents talk, not \emph{whether they reason together}, and say nothing about whose perspectives are in the room to be reasoned about. Human deliberation research has found that procedural and substantive quality can diverge~\cite{knoblochHowDeliberativeExperiences2022,baccaroSmallDifferencesThat2016} and recent work on LLM-as-judge evaluation has independently documented a ``shared illusion'' in which LLM judges agree on surface heuristics while missing substantive quality~\cite{songIllusionConsensusSurface2026}. The simulation of judgment literature has coined \emph{epistemia} for the illusion of knowledge when plausibility replaces verification~\cite{loruSimulationJudgmentLLMs2025}, and a recent roadmap for evaluating moral competence in LLMs identifies the \emph{facsimile problem} of ``models imitating reasoning without genuine understanding'' as a core challenge~\cite{haasRoadmapEvaluatingMoral2026}. We argue that there is a similar risk for LLMs deployed in multi-agent deliberation: the risk that procedural appearance will be mistaken for epistemic substance and that diversity-blind evaluation will mistake convergence among the already-similar for integration across difference.

This paper makes three contributions. 
First, we argue that pluralistic reasoning problems constitute an evaluation gap that neither verifiable-task benchmarks nor procedural discourse metrics fully address. The former do not generalise to the class, the latter capture how agents talk rather than whether they reason together. 
Second, we assemble a three-dimensional necessary-condition test from deliberative practice and validated measurement instruments, spanning procedural quality, outcome quality, and diversity conditions. The Deliberative Reason Index (DRI)~\cite{niemeyerHowDeliberationHappens2024}, a group-level relational measure of intersubjective consistency validated across nineteen citizen-assembly cases, supplies the outcome dimension and, through its response vectors, the basis for diversity assessment. While recent work administered it to individual LLMs~\cite{kreiaumbelinoEmergentUnderstandingHumanAI2025}, we put it to use at the multi-agent group level. Passing all three dimensions does not establish deliberative capacity, but failing any of them precludes it, making the test a relevant constraint on deployment claims.
Third, we draw on empirical evidence from our ongoing research programme on simulating and evaluating LLM deliberation. Across 1{,}980 five-agent runs on 12 citizen-assembly topics and 11 frontier model configurations, the evaluation reveals that LLM groups achieve human-level scores on process quality but produce only small, topic-dependent DRI gains concentrated on tractable rather than ethically contested questions. They begin with roughly one-third of human starting diversity, and invert human perspective convergence dynamics. Our findings suggest that LLM collective reasoning on pluralistic reasoning problems is \emph{procedurally excellent but epistemically shallow}.

We thus draw a line at the distinction between LLMs as \emph{tools} (extending human reasoning, with humans retaining epistemic authority) and LLMs as \emph{epistemic agents} (producing reasoning contributions treated as carrying independent epistemic weight)~\cite{hauswaldArtificialEpistemicAuthorities2025,freimanMakingSenseConceptual2023}. The evidence we review is not a blanket case against AI in deliberation, it supports a targeted argument against a specific assumption: that current LLMs can autonomously integrate pluralistic perspectives into collective reasoning in a reliable way. Several prominent deployments make this assumption implicitly~\cite{rountreeCaseUsingGenerative2026, tesslerAICanHelp2024,fulayEmptyChairUsing2025,fishGenerativeSocialChoice2024}. Our argument suggests these deployments rest on a justificatory gap between the deliberative weight their outputs are considered to have and the epistemic shortcomings our evaluation demonstrates.

We have tested the argument empirically in the political deliberation case because that is where validated DRI instruments exist. Conceptually, the point likely transfers to domains characterized by pluralism and non-verifiability, and adaptation to such domains is a priority for future work.


\section{Evaluating Collective Reasoning}

\subsection{What current benchmarks measure}

Current benchmarks for collective reasoning in multi-agent LLM systems share a defining property: they evaluate against \emph{verifiable solutions}. Whether the task revolves around factual reasoning~\cite{duImprovingFactualityReasoning2023}, logical reasoning~\cite{liangEncouragingDivergentThinking2024,wuCanLLMAgents2025}, multi-agent coordination~\cite{anneHarnessingLanguageCoordination2025,agasheLLMCoordinationEvaluatingAnalyzing2025}, or strategic reasoning in game-like environments~\cite{cipolina-kunGameReasoningArena2025,agarwalWOLFWerewolfbasedObservations2025}, the common structure is that ground truth exists and can be externally checked, or that success has a mathematical or game-theoretic definition. The same property is an important condition for the post-training paradigm that drives current reasoning progress: reinforcement learning with verifiable rewards~\cite{deepseek-aiDeepSeekR1IncentivizingReasoning2025,openaiOpenAIO1System2024}.

Performance on these tasks establishes LLM capacity for a specific kind of reasoning, where reasoning is about finding an answer that can, in principle, be checked. In these situations, multi-agent LLM systems often perform by validating and improving upon the answer of one of the agents~\cite{liangEncouragingDivergentThinking2024}. Pluralistic reasoning problems do not offer an ideal solution that can be verified, making the paradigm of iterative improvement much harder to employ.

\subsection{Pluralistic reasoning}

A substantial class of real-world collective reasoning problems lies outside this scope. Political deliberation on climate policy, healthcare reform, migration, or bioethics has no verifiable answer: perspectives legitimately diverge based on underlying values and framings, and decision quality depends on how well those differences are considered and integrated~\cite{niemeyerEndsDeliberationMetaconsensus2007,dryzekReconcilingPluralismConsensus2006,mouffeDeliberativeDemocracyAgonistic1999}. The same structure characterizes organizational decisions under value conflict, stakeholder negotiations, participatory design, or ethics consultations on emerging technologies. In deliberative-democratic terms, the goal in these settings is not to find a consensus but to construct a \emph{meta-consensus}, i.e. a shared understanding of how relevant considerations map to preferences in order to make more informed and consistent decisions~\cite{niemeyerEndsDeliberationMetaconsensus2007}. Meta-consensus is distinct from agreement on conclusions as participants can disagree sharply about what should be done while sharing an understanding of why they disagree.

Prominent AI-for-democracy deployments target precisely these problems, including AI-mediated consensus-finding~\cite{tesslerAICanHelp2024}, representation of missing perspectives~\cite{fulayEmptyChairUsing2025,zhuCanAITruly2025}, and deliberation simulation~\cite{rountreeCaseUsingGenerative2026}. Similar evaluation challenges arise across many kinds of human--AI decision making on pluralistic reasoning problems~\cite{maHumanAIDeliberationDesign2025,summerfieldImpactAdvancedAI2025}.

The RLVR literature itself acknowledges that verifiable-task techniques do not reach this class: moral reasoning ``typically admit[s] multiple valid answers that reflect different ethical frameworks and value systems, in stark contrast to mathematical and coding problems, which usually have only one objectively correct solution''~\cite{zhangDoesLLMAlignment2026a} and it is being acknowledged that ``success has thus far been largely confined to the mathematical and programming domains with clear and automatically checkable outcomes'', leaving the extension of these techniques to open-ended, value-laden problems an open research question~\cite{zhangExtendingRLVROpenEnded2026a}. We take this as the field's own diagnosis that performance on verifiable tasks does not guarantee capabilities on pluralistic tasks, and there is no principled reason to expect the transfer.

\subsection{Process quality is insufficient}
\label{sec:procedural}

An easy fall back option in situations where verifiable outcome metrics do not apply are procedural measures. These look at the process of coming up with a solution to evaluate the collective reasoning process and infer outcome quality by arguing that high process quality guarantees high outcome quality. In deliberative practice, process quality can be assessed by the Discourse Quality Index (DQI)~\cite{steenbergenMeasuringPoliticalDeliberation2003}. It was developed for human parliamentary deliberation and scores utterances on justification, respect, constructive politics, and engagement with opposing positions. AQuA~\cite{behrendtAQuACombiningExperts2024} automates DQI scoring using an ensemble of adapter models trained on and validated against human coding. These measures capture real dimensions of discourse quality, but they capture \emph{how} agents talk, not \emph{whether they reason together}.

Human deliberation research has documented that procedural and substantive quality can diverge, meaning that well-facilitated, respectful deliberations can fail to produce epistemic gains~\cite{knoblochHowDeliberativeExperiences2022,baccaroSmallDifferencesThat2016}. The same structural concern has been articulated for LLM evaluation more broadly. The Nature roadmap for moral competence in LLMs identifies the \emph{facsimile problem} of models imitating reasoning without engaging it as a foundational challenge~\cite{haasRoadmapEvaluatingMoral2026}. Parallel diagnoses appear in LLM-as-judge research where the ``evaluation illusion'' is described as a phenomenon where evaluators converge on surface heuristics instead of substantive quality ~\cite{songIllusionConsensusSurface2026}) and in studies of LLM judgment where ``the illusion of knowledge emerging when plausibility replaces verification'' has been identified as a key failure mode for reliable LLM judgements~\cite{loruSimulationJudgmentLLMs2025}).

For multi-agent LLM deliberation, this poses a significant risk. If LLM systems produce procedurally high-quality discourse on pluralistic problems while failing to integrate the underlying reasoning, procedural metrics alone cannot distinguish genuine deliberative capacity from its surface imitation. Deliberative LLM agents can then easily be mistaken for epistemic authorities where they actually just mimic deliberative talk without engaging with the underlying reasoning on the issue at hand. 

\subsection{Diversity}
\label{sec:diversity}

Procedural and outcome metrics together evaluate the deliberation \emph{among the participants who are there}, they say nothing about whether those participants span the perspectives the topic demands. For human deliberation this question is typically handled through stratified random selection at the recruitment stage. This diversity is not only meant to ensure representativeness of the mini-public but also, and maybe even more importantly, to ensure cognitively diverse groups. This diversity is what makes deliberation epistemically valuable as the integration of diverse perspectives is considered to help find the best possible solution to a given issue~\cite{landemoreDeliberationCognitiveDiversity2013}.

For multi-agent LLM systems random selection is not an option. Models trained on overlapping corpora start from substantially more similar priors than randomly selected human participants. A growing literature documents that, even when prompted to take specific personas, LLM agents homogenize during debate~\cite{taubenfeldSystematicBiasesLLM2024}, that co-authoring with LLMs shifts human expression toward homogenized linguistic and cultural patterns~\cite{souratiHomogenizingEffectLarge2026}, and that LLM-based social simulation outcomes tend towards unanimity and utopian bias while human data exhibits genuine plurality~\cite{bianSocialSimulationsLarge2025,chenRealworldHumanBehavior2026}. While architectural variation across model families produces some minor differences it therefore does not produce the kind of perspective heterogeneity that human deliberation presupposes.

The implication for evaluation is direct. A multi-agent LLM deliberation that scores well on procedural quality and shows outcome convergence has not necessarily \emph{integrated diverse perspectives}, it may have aligned positions that were never meaningfully apart. Without an explicit diversity measurement, the procedural and outcome dimensions cannot distinguish integration across difference from convergence among the already-similar. This is what motivates treating diversity as a co-equal evaluation dimension rather than a recruitment assumption.


\section{Evaluation Framework and Setup}
\label{sec:framework}

\subsection{From critique to measurement}

Section \ref{sec:procedural} argued that procedural metrics alone leave the evaluation dimensions of epistemic outcome and diversity of perspectives uncovered. This section introduces measurement instruments for both deliberative reasoning and perspective diversity, drawing on instruments validated in deliberative-democracy research. We treat the resulting three dimensions as jointly necessary but individually insufficient. High procedural quality without outcome gains indicates form without substance, outcome gains without diversity indicate convergence among the already-similar rather than integration across difference, and diversity without procedural quality or outcome gains is parallel monologue rather than deliberation. The framework therefore constitutes a \emph{necessary-condition test}. A system that fails on any of the above dimensions cannot be claimed to deliberate, while passing all three is evidence for deliberative capacity but not yet proof of it.

The remainder of this section introduces the outcome measure (\S\ref{sec:dri}), shows how the same instrument supplies a diversity measure (\S\ref{sec:diversity-measurement}), and describes the experimental setup that applies all three to multi-agent LLM deliberation (\S\ref{sec:setup}).

\subsection{The Deliberative Reason Index}
\label{sec:dri}

Following Niemeyer, Veri and colleagues, we operationalise the outcome quality of a deliberation with \emph{meta-consensus}. Meta-consensus refers to the shared understanding of how relevant considerations (i.e. values, perspectives, arguments and opinions) map to preferences (i.e. practical solutions) and does not require participants to agree on their conclusions~\cite{niemeyerEndsDeliberationMetaconsensus2007, dryzekReconcilingPluralismConsensus2006}. The construct is therefore pluralism-preserving by design. Two participants can disagree sharply on a policy while exhibiting high meta-consensus if they share an understanding of how the relevant considerations translate into preferences.

The Deliberative Reason Index (DRI) operationalises meta-consensus as \emph{intersubjective consistency} between participants' considerations and preferences~\cite{niemeyerDeliberativeReasonIndex2022, niemeyerHowDeliberationHappens2024}. Participants rate a set of consideration statements (typically 20--40 items drawn from public discourse on the topic) on a Likert scale and rank a smaller set of policy preferences, both before and after a deliberative intervention. For each pair of participants, Spearman correlations are computed separately across considerations and across preferences. Intersubjective consistency is the similarity of these two correlations. High consistency means that when two participants share or diverge in their reasoning about considerations, this is reflected proportionally in their shared or divergent preferences. Group-level DRI aggregates across all pairs:
\begin{equation}
\mathrm{DRI} = 1 - \frac{2}{n_p}\sum_{(i,j)\in\mathcal{P}}\left|\rho_s(C_i, C_j) - \rho_s(P_i, P_j)\right|,
\end{equation}
where $C_i$ and $P_i$ are participant $i$'s consideration and preference vectors, $\rho_s$ is Spearman correlation, and $n_p = n(n-1)/2$ is the number of unordered pairs. The absolute value prevents positive and negative pairwise discrepancies from cancelling, and the $2/n_p$ normaliser places DRI in $[-1, 1]$, with $1$ at perfect intersubjective consistency. Comparing pre- and post-deliberation DRI scores allows us to understand the impact of a deliberative intervention on the collective reasoning of the group.

The property that makes DRI suitable for our framework is that it \emph{distinguishes genuine epistemic gains from surface agreement}. Groups can produce consensus without meta-consensus (agreement on conclusions without shared reasoning). They can also produce high procedural quality without meta-consensus (well-mannered discussion that fails to integrate perspectives), and they can produce meta-consensus without consensus (shared reasoning amid persistent disagreement). DRI separates these cases cleanly, which is exactly the separation that procedural metrics alone cannot make. The measure is additionally content-agnostic as DRI requires no external judgment about what conclusions are good or which considerations matter and it is group-level relational as it captures a property of the group's reasoning structure that no individual-level measure can. DRI has been validated across nineteen citizen-assembly-scale deliberations on topics including climate policy, healthcare, constitutional reform, and bioethics, where it reliably increases over successful deliberations and is robust to information provision, group conformity, and expert framing~\cite{niemeyerHowDeliberationHappens2024}.

\subsection{The diversity metric}
\label{sec:diversity-measurement}

The diversity dimension is operationalised by vectorising the participants' DRI responses. For a group of $n$ participants with standardised response vectors $x_i \in \mathbb{R}^d$ spanning all consideration and preference items, we compute mean pairwise Euclidean distance:
\begin{equation}
D(x_1, \ldots, x_n) = \frac{2}{n(n-1)} \sum_{1 \le i < j \le n} \|x_i - x_j\|_2.
\end{equation}
Pre-deliberation diversity measures whether the group meets Landemore's requirement that deliberation engages genuinely different starting positions~\cite{landemoreDeliberationCognitiveDiversity2013}. Pre-to-post change measures the \emph{convergence dynamic}. Human deliberation typically reduces dispersion as diverse perspectives synthesise into shared understanding revolving around several clusters of opinions.

\subsection{Experimental setup}
\label{sec:setup}

A recent benchmark study applies the three-dimensional framework to multi-agent LLM deliberation~\cite{flechtnerProceduralParityOutcome2026}. The relevant aspects of its setup are summarised below. Full statistical specifications and robustness checks appear in the cited work. The deliberation and survey prompts it uses are presented in Appendix~\ref{app:prompts}. The findings reviewed in \S\ref{sec:findings} draw on this study, supplemented with a persona-prompting pilot introduced below and an individual-level study assessing LLM alignment with human DRI answers~\cite{kreiaumbelinoEmergentUnderstandingHumanAI2025}.

Each deliberation in the study involves five LLM agents engaging on a single topic over two rounds, with speaking order randomised within rounds, the full transcript passed forward between turns, and decoding temperature fixed at zero for reproducibility. Each agent completes the DRI survey before and after deliberation, rating the topic's consideration statements and ranking its policy preferences. Three prompting regimes are compared: a survey-only baseline that establishes same-system noise by moving from the pre- to the post-survey with no deliberation in between, a basic prompt that relies on the model's internal understanding of the term, and a normative prompt that explicitly invokes the deliberative norms underpinning DQI and DRI. The full grid crosses eleven model configurations, twelve topics, and three treatments with five replicates each, for 1{,}980 runs.

The eleven model configurations span five frontier families (GPT-5.1, Gemini-3-Pro-Preview, Claude Opus 4.5, DeepSeek-V3.2-Exp, Kimi-K2-Thinking), with reasoning-enabled and standard variants where available, plus two mixed-family ensembles that draw each agent from a different family. The twelve topics are drawn from citizen assemblies with validated DRI instruments and matched human pre/post survey data, spanning climate policy, healthcare, governance, bioethics, and urban planning.

Two human reference distributions anchor the analysis. Procedural quality is benchmarked against the Europolis deliberative poll~\cite{gerberDeliberativeAbilitiesInfluence2018} ($N = 910$), the same corpus used to validate AQuA. Outcome quality and diversity are benchmarked against the human survey data from the twelve citizen assemblies ($N = 407$). Statistical inference is topic-aware throughout: the study reports topic-blocked permutation tests, topic-resampled bootstrap confidence intervals, and Holm correction for multiple comparisons, with a hierarchical mixed-effects model (topic random intercepts, model random slopes) as a complementary specification. This reflects substantial topic-level clustering (ICC $\approx 0.11$ for $\Delta$DRI) against limited topic-level replication ($n = 12$).

An additional pilot tests whether the homogeneity observed across the main study reflects identical agent instructions rather than properties of the models themselves. Empirically grounded personas are constructed by clustering real citizen-assembly survey responses with $k=5$ per topic (k-means on standardised consideration response vectors), then translating each cluster's defining considerations into a natural-language value profile delivered as the agent's system prompt. We cluster on considerations alone, not preferences, so that persona-prompted agents derive their preferences from the considerations they hold rather than anchoring on a fixed policy ranking, mirroring the direction of reasoning deliberation is meant to elicit. The pilot is run on three topics across two models ($N=60$ deliberations). Full clustering details, persona-prompt templates, and example personas appear in Appendix~\ref{app:personas}.


\section{Evidence: The Three-Dimensional Framework Applied}
\label{sec:findings}

This section synthesizes evidence from the benchmark study of multi-agent LLM deliberation~\cite{flechtnerProceduralParityOutcome2026}, combined with an individual-level study of 54 LLMs against 526 human participants across 24 citizen-assembly cases~\cite{kreiaumbelinoEmergentUnderstandingHumanAI2025} and the persona-prompting pilot described above. The analysis will follow the three-dimensional framework introduced in \S\ref{sec:framework}. Here we summarise only the dimensions the argument turns on, reported against human reference distributions on the same instruments. Full model-by-model and topic-by-topic breakdowns and the protocol-level robustness pilots are reported in the benchmark study, where simulation and analysis code is provided as a supplement~\cite{flechtnerProceduralParityOutcome2026}.

\begin{table}[t]
\centering
\setlength{\tabcolsep}{2pt}
{\small
\begin{tabular}{lcccc}
\toprule
\multicolumn{5}{l}{\emph{Outcome} --- pooled $\Delta$DRI contrasts} \\
Contrast & ATE & 95\% CI & $p$ & $p_{\text{Holm}}$ \\
\midrule
Normative vs.\ None & $0.029$ & $[-0.004, 0.076]$ & $0.005$ & $0.015$ \\
Basic vs.\ None     & $0.019$ & $[-0.011, 0.057]$ & $0.079$ & $0.158$ \\
Human reference     & $0.099$ & $[0.026, 0.196]$  & ---     & ---     \\
\midrule
\multicolumn{5}{l}{\emph{Procedural} --- mean AQuA (0--4)} \\
Group & Mean & SD & $N$ & \\
\midrule
LLM (Basic $+$ Norm.) & $2.939$ & $0.12$--$0.13$ & $1{,}320$ & \\
Human (Europolis)     & $2.980$ & $0.431$        & $910$     & \\
\midrule
\multicolumn{5}{l}{\emph{Diversity} --- mean pairwise distance} \\
Group & Pre & Post & $\Delta$ & \\
\midrule
Human (topic means) & $18.78$ & $17.57$ & $-1.21$ & \\
LLM (Normative)     & $6.51$  & $6.77$  & $+0.26$ & \\
LLM (Basic)         & $6.56$  & $6.93$  & $+0.37$ & \\
\bottomrule
\end{tabular}}
\caption{Three-dimensional evidence synthesised from the benchmark
study~\cite{flechtnerProceduralParityOutcome2026}. Outcome CIs are
topic-resampled bootstrap intervals, $p$-values are from topic-blocked
permutation tests. AQuA dispersion is reported as SD,
the LLM--human AQuA contrast gives $p_{\text{Holm}}=0.052$. Diversity is mean
pairwise Euclidean distance on standardised response vectors.}
\label{tab:synthesis}
\end{table}

\subsection{Procedural quality}
\label{sec:findings-procedural}

On procedural quality, LLM groups clear the deliberative bar AQuA operationalises. As AQuA scores discourse transcripts, it is only computed on the 1{,}320 sessions that involve deliberation (the basic and normative prompting regimes), not on the 660 survey-only baseline runs that make up the remainder of the 1{,}980-run study. Across those 1{,}320 sessions, mean AQuA (2.939) sits just below the Europolis human reference (2.980), a gap that does not reach significance after Holm correction ($p_{\text{Holm}} = 0.052$)~\cite{flechtnerProceduralParityOutcome2026}. We read this as procedural competence at human-comparable levels. LLM discourse also shows substantially lower variance (SD 0.12--0.13 vs.\ 0.43 for humans). Decomposition across the twenty AQuA dimensions shows that LLM discourse matches or approximates humans on justification, respect, engagement, and constructive politics, with negligible rates of disrespect, vulgarity, or sarcasm. By the procedural standard alone, LLM deliberation looks competent, so the framework's first dimension is passed.

\subsection{Outcome quality is small, topic-dependent, and concentrated on tractable problems}
\label{sec:findings-outcome}

On outcome quality measured by DRI, the picture is very different. Under explicit normative prompting, multi-agent LLM deliberation produces only small absolute improvements in intersubjective consistency ($\Delta\mathrm{DRI} \approx 0.029$, topic-blocked), roughly 30\% of the human reference ($\Delta\mathrm{DRI} \approx 0.099$). Under basic prompting that relies on the model's internal understanding of ``deliberation'', the effect is smaller and not reliably different from zero. The pooled effect under normative prompting is statistically detectable but does not survive Holm correction at the topic level, and turns \emph{negative} on ethically contested topics such as Swiss healthcare reform and Uppsala begging policy~\cite{flechtnerProceduralParityOutcome2026}.

The topic-heterogeneity pattern is itself diagnostic. Effects are largest on concrete, locally-bounded questions where the mapping between considerations and preferences is relatively unambiguous (the Fremantle bridge question shows the largest positive effect, $\approx 0.24$). Effects attenuate or reverse on abstract, value-laden topics where the consideration-preference mapping is itself the contested terrain. LLM deliberation thus works better on the practical questions that are less value sensitive and fails on the contested questions where deliberative reasoning is most needed.

Frontier LLMs, even when prompted with detailed deliberative instructions, therefore do not reliably produce the epistemic effects deliberation is designed to achieve on pluralistic reasoning problems. The capability is neither reliably present nor reliably absent, it is present inconsistently, dependent on topic, and fragile under reasonable corrections for multiple testing. The framework's second dimension is thus not robustly passed.

\subsection{Individual-level corroboration}
\label{sec:findings-individual}

In a separate study, Kreia Umbelino and Veri~(\citeyear{kreiaumbelinoEmergentUnderstandingHumanAI2025}) evaluated 54 off-the-shelf LLMs against 526 post-deliberation human DRI-responses across 24 cases spanning 19 topics. In this setup, each LLM's answers to the DRI survey were compared against the human answers from the corresponding case~\cite{kreiaumbelinoEmergentUnderstandingHumanAI2025}. Humans significantly outperform LLMs on average ($\mu_{\mathrm{LLM}} = 0.18$ vs.\ $\mu_{\mathrm{human}} = 0.34$, $p < 0.0001$), indicating that LLMs deliberative reasoning patterns do not match human patterns on the same issue. At the case level, humans outperform LLMs in 19 of 24 cases. In some of the remaining cases, human post-deliberation DRI is itself anomalously low or affected by expert-stakeholder dynamics, suggesting that LLM parity reflects matched difficulty rather than matched capability. At the model level, humans consistently outperform 23 of the 54 models tested, while the remaining majority perform on par with humans on average across cases.

Model-level variation does not follow clear patterns of scale, recency, or reasoning capability as the latest reasoning-enabled models do not systematically outperform their non-reasoning counterparts on this task, consistent with work documenting trade-offs between task performance and deliberative reasoning in reasoning-tuned models~\cite{zhaoTradeoffsLargeReasoning2025}. These LLMs were tested off-the-shelf, without fine-tuning or access to deliberation transcripts, so the individual-level finding establishes a baseline rather than a ceiling. Within that baseline, the convergence of group-level and individual-level evidence is what bears on the argument here: the outcome-dimension shortfall observed at the group level is not an artifact of the group-level analysis but reflects a property of LLM responses to deliberative reasoning instruments more generally. The individual-level finding has specific consequences for deployments that propose LLMs as representatives of absent human perspectives, which we return to in \S\ref{sec:deployments}.
\subsection{Severe under-diversity and reversed convergence}
\label{sec:findings-diversity}

Human deliberative groups begin with substantial perspective heterogeneity (mean pairwise Euclidean distance $\approx 18.8$ on standardised survey vectors) and \emph{converge} through deliberation ($\Delta \approx -1.21$), consistent with the theoretical expectation that deliberation integrates diverse perspectives into shared understanding~\cite{landemoreDeliberationCognitiveDiversity2013, niemeyerHowDeliberationHappens2024}. LLM groups, across all models and treatments tested, begin at approximately one-third of human starting diversity ($\approx 6.5$) and either remain stable or \emph{diverge} slightly through interaction ($+0.26$ to $+0.37$ under deliberation treatments)~\cite{flechtnerProceduralParityOutcome2026}. Mixed-model ensembles combining different model families show modestly higher starting diversity ($\approx 9.5$) but remain far below human levels and exhibit the same trajectory pattern.

Human deliberation derives much of its epistemic value from bridging genuinely different perspectives~\cite{landemoreDeliberationCognitiveDiversity2013}. Under these circumstances, the integration of diverse considerations \emph{is} the reasoning work. LLM groups begin without the diversity that makes this integration necessary, because they inherit similar values, framings, and argumentative moves from related training distributions~\cite{taubenfeldSystematicBiasesLLM2024, souratiHomogenizingEffectLarge2026}. When LLM groups do show DRI gains, those gains reflect the alignment of already-similar reasoning structures rather than the reconciliation of pluralistic perspectives.

The framework's third dimension fails decisively. And because diversity is constitutive of deliberation's epistemic value, this failure has consequences beyond the diversity dimension itself. Even where outcome quality appears to improve, the improvement does not indicate the integration deliberation is meant to produce.

\subsection{Engineering diversity does not restore the human dynamic}
\label{sec:findings-personas}

\begin{table}[t]
\centering
\setlength{\tabcolsep}{2pt}
{\small
\begin{tabular}{lcc}
\toprule
Measure & LLM (persona) & Human ref. \\
\midrule
Pre-delib.\ diversity                & $27.7$ & $\approx 18.8$ \\
$\Delta$DRI                          & $-0.058$ (n.s.) & $+0.099$ \\
$\Delta$Consideration agreement      & $+0.066$ ($d{=}0.55$) & $+0.077$ \\
$\Delta$Preference agreement         & $-0.008$ (n.s.) & $+0.104$ \\
\bottomrule
\end{tabular}}
\caption{Persona-prompting pilot ($N=60$; 3 topics, 2 models). Engineered
diversity raises starting heterogeneity above the human reference but does not
improve $\Delta$DRI, and inverts the human update pattern: agreement rises on
considerations, not preferences. Human decomposition references
($+0.077$, $+0.104$) are the pilot-topic human baseline. The $\Delta$DRI human
reference is the twelve-assembly mean. Diversity $p<0.001$, consideration
agreement $p=0.039$, preference agreement and $\Delta$DRI n.s.\
($p=0.456$).}
\label{tab:persona}
\end{table}

A natural objection is that observed homogeneity reflects identical agent instructions rather than properties of the models themselves. A persona-prompting pilot tests this directly. Empirically grounded personas are constructed by clustering human DRI survey responses ($k=5$ per topic) and translating cluster-defining considerations into natural-language value profiles delivered as agent system prompts (see \S\ref{sec:setup}, $N=60$ deliberations across three topics and two models).

The manipulation succeeds at the input level. Pre-deliberation diversity rises from $\approx 7.5$ to $\approx 27.7$ ($+20.23$, $p < 0.001$), exceeding the human reference. But $\Delta$DRI does not improve ($\beta = -0.058$, $p = 0.456$). Decomposing the result reveals an inversion of the human pattern. In human assemblies, deliberation produces larger convergence in \emph{preference agreement} ($+0.104$) than in \emph{consideration agreement} ($+0.077$). So participants update preferences to align with the considerations they share. Persona-prompted LLMs reverse this pattern. Consideration agreement rises significantly ($\Delta = +0.066$, $d = 0.55$, $p = 0.039$), while preference agreement does not change ($\Delta = -0.008$, n.s.), so LLMs tend to update their considerations to match their preferences.

The sample size is small ($N=60$) and the pilot is diagnostic rather than confirmatory. But the inversion indicates that LLM groups update on the dimension humans hold stable and fail to update on the dimension humans converge, when engaging different perspectives. Thus, engineering diversity does not produce deliberative dynamics, it changes which component of deliberative reasoning fails.

\subsection{Framework verdict and failure mode}
\label{sec:findings-verdict}

Investigating LLM deliberation with this three-dimensional framework, we find that LLM deliberation is procedurally competent at human-comparable levels, but is inconsistent on the outcome dimension, failing specifically where the deliberative work is most normatively laden. The diversity dimension of LLM deliberation differs from human diversity in a manner that engineering interventions cannot restore. According to the necessary-condition test the framework articulates, current LLM deliberation does not warrant claims to deliberative capacity on pluralistic reasoning problems. We call this combined failure across the epistemic outcome and diversity dimensions the \emph{deliberative deficit}.

Qualitatively, the pattern is consistent across transcripts~\cite{flechtnerProceduralParityOutcome2026}. Agents produce extended, well-justified, respectful exchanges in which they acknowledge each other's contributions, build on shared framings, enumerate balanced considerations, and collaboratively construct elaborate policy frameworks. What is missing is substantive disagreement. Agents rarely defend incompatible positions, rarely push back on each other's fundamental framings, and rarely integrate considerations that were genuinely in tension. Apparent divergence, when it occurs, typically takes the form of different agents elaborating different aspects of a shared position rather than defending opposing views.

Mechanistically, these findings could be explained by LLMs arriving at deliberation with pre-formed shared representations from overlapping training data or similar post-training treatment. They already share the meta-consensus human deliberation has to construct and they share it without the integration work that gives meta-consensus its democratic value. Their reasoning work collapses into collaborative elaboration of an already-shared frame. We call this failure mode \emph{procedurally excellent, epistemically shallow} deliberation. It mirrors the \emph{facsimile problem} identified for individual moral reasoning~\cite{haasRoadmapEvaluatingMoral2026}. Where the facsimile problem identifies imitation without understanding at the individual moral-reasoning level, we identify a structural analogue at the collective reasoning level. Surface performance is real, but deliberative work is absent, and the absence is invisible to the evaluations most likely to be used. For collective reasoning on contested problems, this failure mode is more dangerous than obvious incapacity as it invites unwarranted confidence in deployments that the underlying capability does not support.


\section{The Deliberative Deficit and Its Consequences}
\label{sec:findings-llms}

The findings characterise a deliberative deficit of current LLMs. While they produce discourse that meets procedural standards of deliberation, their internal and collective reasoning patterns often do not align with human reasoning patterns on the same issue. Furthermore, the diversity that gives deliberation its democratic and epistemic value is something that LLMs neither bring to a conversation nor learn to integrate when prompted to perform it. The capacity to reason across and integrate genuinely different perspectives is however, what distinguishes deliberation on pluralistic reasoning problems from collaborative elaboration as tested in collective reasoning benchmarks on verifiable tasks, and it is a capacity LLMs cannot reliably display.

\subsection{Deployments that implicitly assume epistemic agency}
\label{sec:deployments}

Contrary to our findings, several prominent deployment lines rest on the assumption that current LLMs can perform such deliberative reasoning on pluralistic problems. We analyse four classes through the framework's lens. For each we explain what the deployment claims, what assumption that claim rests on, and what evidence in \S\ref{sec:findings} bears on that assumption.

\paragraph{AI representation of missing perspectives.} Deployments that propose LLMs as representatives of underrepresented or absent perspectives~\cite{fulayEmptyChairUsing2025, zhuCanAITruly2025} claim that an LLM can reason \emph{from} a perspective in ways that contribute epistemically to group deliberation. This claim rests on the assumption that LLMs, when prompted with appropriate persona instructions, can contribute the perspective's considerations to the discussion in a way that reflects how a holder of that perspective would actually weigh trade-offs, and, on the other hand, integrate the others' perspectives into its own reasoning, updating its internal reasoning framework accordingly.

The evidence undermines this assumption on two fronts. The individual-level study by Kreia Umbelino and Veri (\citeyear{kreiaumbelinoEmergentUnderstandingHumanAI2025}) shows that off-the-shelf LLMs systematically fall below human DRI alignment with the corresponding human participants in each case (\S\ref{sec:findings-individual}). Such a deployment claims that LLM reasoning aligns with how humans in the relevant case would reason. This is a capability many current systems lack at the individual level. Additionally, the persona-prompting pilot (\S\ref{sec:findings-personas}) shows that LLMs with engineered persona diversity do not mimic human deliberation dynamics. When LLM groups are forced to engage with genuinely different perspectives, they converge on considerations not preferences and leave DRI unchanged. Engineered personas do not produce the integration that the deployment requires for LLMs to genuinely represent a perspective in deliberation.

\paragraph{AI-mediated consensus-finding.} The most influential application in AI-mediated deliberation is probably the ``Habermas Machine''~\cite{tesslerAICanHelp2024}. It reports that LLM-generated consensus statements outperform statements produced by human mediators. Participants prefer the LLM outputs, and the system scales group agreement-finding in ways human mediation cannot.

The validation rests on participant rankings of LLM-generated statements rather than on direct deliberation between participants. The LLM produces a statement, participants rate it, the LLM revises, and convergence emerges from this iteration. While this framework has already been criticized elsewhere~\cite{hernandezAutomatingDeliberationIdea2025}, we argue that our findings can add a further epistemic dimension to the critique. The procedure establishes that the LLM is effective at producing statements that participants rank highly, but successful deliberation is not defined by user satisfaction, but by the epistemic effect it has~\cite{niemeyerEndsDeliberationMetaconsensus2007, estlundEpistemicValueDemocratic2018}. These are not the same outcome, and the framework described in \S\ref{sec:framework} is designed to enable the evaluation of the latter. Meta-consensus is a group-level relational property that emerges from participants reasoning together and integrating each other's reasoning. The acceptability of LLM-summarisations, on the other hand, is a property of the LLM's capacity to find framings that score well across heterogeneous starting points. The Habermas Machine's validation cannot tell whether participants engaged with diverging opinions and reasoned their way to common ground or whether the LLM found a framing acceptable enough to all, creating the appearance of common ground instead of constructing it.

The ranking-based validation has an additional vulnerability. Current LLMs have been shown to struggle to produce coherent rankings that reflect integrated consideration of diverse values~\cite{kreiaumbelinoEmergentUnderstandingHumanAI2025}. The Habermas Machine relies on predicting which framings will rank highly across heterogeneous participants, which is the inverse capability of that. The system's apparent success on ranking-based outcomes is consistent with the failure mode of procedural performance that does not index the epistemic engagement deliberation requires~\cite{oleartWhyAITechnosolutionism2025}. We do not claim the Habermas Machine result is wrong, but that it is underdetermined. A DRI-type evaluation could help distinguish whether the common ground reflects participants integrating their perspectives across differences or the LLM finding statements acceptable enough to satisfy users.

\paragraph{Democratic simulation and digital representatives.} Deployments that simulate citizen deliberation at scale or propose LLM agents as digital representatives of citizens~\cite{jarrett2025languageagentsdigitalrepresentatives, novelliReplicaOurDemocracies2025, rountreeCaseUsingGenerative2026, lowHabermoltDelegatingDeliberation2026} are based on the assumption that LLM agents reason in ways that approximate how citizens would. The presented diversity findings directly challenge this assumption as LLM groups begin at roughly one-third of human starting diversity and either remain stable or diverge through interaction (\S\ref{sec:findings-diversity}), which is the opposite of the human convergence pattern. As LLMs begin without and do not develop the perspective heterogeneity that constitutes deliberation's epistemic value, simulated democratic deliberation differs from real democratic deliberation not in fidelity but in kind. An LLM-simulated democracy is therefore not a lossy approximation of a real one but a structurally different object with the tendency to shift opinions towards baseline perspectives during deliberation~\cite{taubenfeldSystematicBiasesLLM2024}. 

\paragraph{Generative social choice.} Approaches that use LLMs to generate statements representing ``cohesive coalitions'' and predict participant preferences over novel statements~\cite{fishGenerativeSocialChoice2024} provide theoretical guarantees of representational fairness based on the condition that LLMs have the capacity to capture the structure of diverse preferences. Their framework assumes LLMs can predict participant preferences over novel statements. The evidence presented in \S\ref{sec:findings} suggests this capacity is weak, especially on contested topics where the approach is most valuable.

The goals of these deployment lines are legitimate, and several reflect serious attention to representational fairness and procedural design. But none of the proposals question the underlying assumption that current LLMs can autonomously integrate pluralistic perspectives.

\subsection{Generalisation beyond political deliberation}
\label{sec:generalization}

The conceptual argument extends to further domains characterised by pluralism and non-verifiability, such as ethics consultations, organisational strategy under value conflict, stakeholder negotiations, participatory design, or contested resource allocation. LLM deployments in these domains face the same evaluation gap~\cite{maHumanAIDeliberationDesign2025, summerfieldImpactAdvancedAI2025}. DRI as currently validated is specific to political deliberation with domain-calibrated survey instruments, and extension to other domains requires adapted measurement. The conceptual machinery of meta-consensus as the evaluation target, group-level relational measurement, and joint reporting of procedure- and outcome-diversity, would transfer directly. Adapting the instruments is itself a research priority.

In day-to-day usage, users routinely interact with LLMs as if they were deliberative entities. When arguing with them, trying to persuade them, asking them to weigh contested considerations and arrive at consistent positions, users treat the LLMs' responses as if they reflect integrated reasoning across perspectives. The LLMs' fluency, their apparent reasonableness and willingness to engage with counter-arguments are all procedural markers of deliberation. What is absent is the underlying capacity those markers normally index in human interaction. The LLM that ``considers both sides'' has not constructed a shared reasoning framework across them, the LLM that ``updates its position'' has not integrated a perspective it did not already approximate, and the LLM that ``finds common ground'' has not bridged perspectives it never genuinely held in tension.

Benchmarks dominated by verifiable collective reasoning tasks can easily be misinterpreted to transfer over to similar tasks pluralistic, non-verifiable problems. However, the two relate to fundamentally different domains and the presented evidence suggests that performance does not necessarily transfer from one to the other. Recent work on pluralistic alignment~\cite{sorensenPositionRoadmapPluralistic2024, PeterDecentralizing2025}, moral reasoning benchmarks that formalise procedural and pluralistic evaluation~\cite{chiuMoReBenchEvaluatingProcedural2025}, and the conceptualisation of the facsimile problem in moral competence~\cite{haasRoadmapEvaluatingMoral2026} have begun to address this gap. What has been missing is a group-level relational measure that captures the collective dimension of pluralistic reasoning. Applying DRI at the group level, as the discussed benchmark study does~\cite{flechtnerProceduralParityOutcome2026}, supplies one.

\subsection{Legitimate tool uses}
\label{sec:tools}

The appropriate response to the evidence is role constraint. We argue that LLMs can legitimately support collective reasoning on pluralistic problems in ways that do not require them to function as autonomous epistemic agents, but should not be treated as such until the epistemic gap is closed.

Based on work on the epistemic role of AI~\cite{hauswaldArtificialEpistemicAuthorities2025, alvaradoAIEpistemicTechnology2023}, and the conceptual limits of attributing normative commitments to AI systems~\cite{freimanMakingSenseConceptual2023, ferrarioExpertsAuthoritiesStrange2024}, we distinguish two roles. An \emph{epistemic agent} is a participant whose reasoning contributions are treated as carrying independent epistemic weight, e.g. when contributions enter the group's reasoning \emph{as reasoning}, not merely as inputs to be processed by humans who retain authority over their interpretation. A \emph{tool} instead extends or supports human reasoning while humans retain epistemic authority over what counts as a good reason and whether the group has reasoned well together. The distinction is not whether the system is sophisticated, it is whether the human consumers of its outputs treat those outputs as already-reasoned positions in the deliberation or as inputs to their own deliberation.

For tool use, what matters is the procedural quality of LLM output and its effect on human reasoning. For the use as an epistemic agent, the LLM itself must be evaluated with much more scrutiny. The evidence reviewed in \S\ref{sec:findings} establishes that LLMs pass the procedural test, are inconsistent on outcome, and fail on diversity in ways engineering interventions do not repair. According to the presented evidence, the deliberative weight that LLM outputs carry when treated as epistemic agents is not currently justified by the systems that produce their outputs.

The tool/agent distinction also suggests deployments that are compatible with current LLM capabilities. Information synthesis for participants, translation and accessibility accommodations, logistical scaffolding for deliberation organisation, individual reflection prompts that surface considerations a participant might not have weighed, and moderation support on procedural violations are all roles in which an LLM extends human deliberative capacity without being required to autonomously integrate pluralistic perspectives. In these cases, the human participants still have to do the deliberative work, while the LLM extends their capacity to do it. The diagnostic test is whether users are inclined to engage with LLMs as conversation partners instead of output generators. As soon as users aim to argue with or convince an LLM on a pluralistic reasoning problem, they are engaging with the model in a way that assumes that it is an epistemic agent that has the capacity and motivation to aim for a more consistent reasoning structure. However, the evidence reviewed here suggests current LLMs currently lack this capacity.

\subsection{What follows for evaluation practice}

If a proposed AI-application includes functions of an epistemic agent and revolves around pluralistic reasoning problems, we argue that its evaluation should include at least the dimensions of procedural quality, deliberative reasoning quality and diversity. Ideally, these dimensions should be reported against human reference distributions on the same questions. Particular attention should be given to the relationships between the dimensions as outcome gains without diversity have to be interpreted differently than outcome gains with diversity, and procedural quality alone cannot justify inferences about either of the other two dimensions.

\subsection{Limitations}
\label{sec:limitations}

The proposed evaluation framework and the presented argument around it come with their own set of limitations, constraining how its conclusions should be read.

\paragraph{Computational instruments for a political question.}
The presented framework judges the LLMs' capacities for democratic deliberation using computational instruments. Any instrument that formalises deliberation cannot register what resists formalisation, and a system optimised against it can pass while missing exactly that residue. DRI in particular reads meta-consensus off a correlation structure but is silent on power, recognition, and the non-propositional modes of communication that a substantial strand of democratic theory treats as constitutive~\cite{youngInclusionDemocracy2000}. The diversity metric captures dispersion in a rated response space, not the social diversity it proxies. Under deployment, a metric that hardens into a gate reshapes what counts as good deliberation, displacing the contestation over what deliberation is for~\cite{mouffeDeliberativeDemocracyAgonistic1999} and reproducing the technosolutionism in the evaluation apparatus that is often warned against~\cite{oleartWhyAITechnosolutionism2025}.

\paragraph{Behavioural signature versus underlying process.} DRI assesses how consistently a group maps particular consideration statements to policy preferences. It therefore does not measure the deliberative reasoning as an underlying cognitive process directly but only a behavioural signature of it. We cannot distinguish whether LLMs that show DRI gains under deliberation are genuinely integrating perspectives or just mimic the fingerprint of such an integration to produce coherent-looking responses. The persona-prompting pilot's inversion finding is consistent with the second reading as LLMs are clearly more inclined to adjust their values to policy preferences than the other way around. Considering this limitation, we frame the evidence as a necessary-condition test rather than a sufficient one for this reason. If LLM systems fail to produce even the behavioural signature of deliberation under favourable conditions, claims that they deliberate in any richer epistemic sense are not supported. Whether they could be made to deliberate under different training or prompting regimes is an open question the methodology cannot currently settle.

\paragraph{Statistical power and replication.} The benchmark uses twelve citizen-assembly topics with validated DRI instruments. Topic-level effects do not survive Holm correction, and the persona-prompting pilot is run on three topics across two models with $N=60$ deliberations. The findings are robust at the pooled level and to hierarchical mixed-effects modelling, but topic-specific claims should be read as suggestive rather than confirmatory. Replication on additional topic sets, particularly outside Western and especially Australian deliberative contexts, would strengthen what the present evidence can support.

\paragraph{Framework commitments.} The three-dimensional framework operationalises deliberation through procedural discourse quality, DRI outcome quality, and perspective diversity. These are coherent within deliberative-democratic theory but reflect a specific tradition. Especially aggregative democratic frameworks might evaluate AI applications in democratic settings differently. An aggregative theorist would likely care more about distortions in preference aggregation than about the quality of reasoning that precedes it. The argument applies on deliberative-democratic terms, whether it survives translation into other democratic traditions is a separate question the paper does not address.

\paragraph{Off-the-shelf systems with standard prompting.} The benchmark evaluates vanilla frontier LLMs. Fine-tuning specifically for deliberative reasoning, retrieval augmentation, hybrid human-AI protocols, and persona taking are not tested. The claims should be read in light of this limitation. Where the paper says current LLMs cannot reliably perform deliberative reasoning on pluralistic problems, the qualifier is 'current off-the-shelf systems under reasonable prompting'.


\section{Conclusion}
\label{sec:conclusion}

The deliberative deficit we have characterised points out how current LLMs struggle to reason across pluralistic perspectives in an integrative way. This deficit is hard to detect from procedural evaluation or verifiable-task benchmarks alone. With the latter driving current reasoning progress successfully in relevant domains, we warn against overstretching the generalization to cases of pluralistic reasoning problems. Systems that produce fluent, respectful, reason-giving discourse and converge on common framings look like they are deliberating, and the institutions that deploy such systems in democratic settings are licensed by that appearance to claim outputs as deliberatively produced. The evidence reviewed here suggests the appearance is not currently earned.

The three-dimensional test we apply is intended to help discover this kind of misinterpretation. It defines three dimensions on which deliberative capacity has to be demonstrated and the necessary-condition logic it articulates further constrains when developers can credibly claim that an LLM application is doing deliberative work. The empirical evidence we draw on shows that frontier LLMs do not meet this standard. Our conclusion is open to revision under different training paradigms, hybrid human-AI protocols, or evaluation regimes we do not test here. What it does establish, on present evidence, is that the procedural fluency LLMs reliably produce should not be used single-handedly to infer deliberative substance, and that interventions specifically designed to fix the diversity dimension do not produce the human deliberative dynamic but change which component fails.

For future work, three directions follow. First, the framework should be extended to test non-political domains where pluralistic reasoning matters but where validated instruments do not yet exist. Ethics consultations, organisational strategy under value conflict, and stakeholder negotiations in contested resource allocation all carry the same evaluation gap, and the conceptual machinery of the test transfers. Second, it should be investigated whether LLMs could be trained to construct meta-consensus rather than approximate it from overlapping priors. A training paradigm that rewards relational coherence between considerations and preferences across multiple perspectives could potentially lead to significant performance gains across domains. Third, the tool/agent distinction we have drawn is operational, but its translation into evaluation regimes, deployment standards, and procurement requirements for AI in democratic contexts is institutional work the present paper does not undertake. 

The broader implication is for how AI capacity is evaluated across the field. Pluralistic, non-verifiable collective reasoning is a substantial class of real-world problems on which LLMs are increasingly deployed, and the evaluation apparatus the field has built does not reach it. The three-dimensional test we apply could help to close that gap.

\appendix
\section*{Appendix}

\section{Prompts}
\label{app:prompts}

The benchmark study administers the DRI survey to each agent before and after deliberation, and prompts deliberation under two regimes (basic and normative). The prompts are reproduced here for self-containment. Bracketed slots are filled in per topic and per agent.

\subsection{DRI survey prompts}

\begin{promptbox}{Pre-Deliberation DRI Survey Prompt}
Topic: [ISSUE\_TOPIC]

Indicate your agreement to each consideration on a scale of [MIN]-[MAX]
with [MIN] being strongly disagree and [MAX] being strongly agree:

C1. [CONSIDERATION\_1\_TEXT]
   Rating ([MIN]-[MAX]):
C2. [CONSIDERATION\_2\_TEXT]
   Rating ([MIN]-[MAX]):
...

Rank the following [N] preferences (1 being highest priority):

P1. [PREFERENCE\_1\_TEXT]
   Rank:
P2. [PREFERENCE\_2\_TEXT]
   Rank:
...

Please provide your responses in exactly the following format:
Considerations:
C1: <rating>
C2: <rating>
...

Preferences:
P1: <rank>
P2: <rank>
...

Do not include any other text than the format above.
\end{promptbox}

\begin{promptbox}{Post-Deliberation DRI Survey Prompt}
Topic: [ISSUE\_TOPIC]

Context from previous discussion:
---
[DELIBERATION\_TRANSCRIPT]
---

Here were your previous answers:

Considerations Ratings:
- C1: [PREVIOUS\_RATING\_1]
- C2: [PREVIOUS\_RATING\_2]
...

Preference Rankings:
- P1: [PREVIOUS\_RANK\_1]
- P2: [PREVIOUS\_RANK\_2]
...

Please review the discussion context and update your ratings and rankings
below based on the deliberation.

[Same format as pre-survey follows]
\end{promptbox}

\subsection{Deliberation prompts}

\begin{promptbox}{Basic Deliberation Prompt}
You're taking part in a citizen's assembly on the topic: [TOPIC]
Deliberate with the other participants.
\end{promptbox}

\begin{promptbox}{Normative Deliberation Prompt}
You're taking part in a citizen's assembly on the topic: [TOPIC]
Deliberate with the other participants.

Be respectful, reasoned, and authentic. Do not use force in your language.
Express your viewpoints and give reasons. Orient your arguments and
viewpoints towards the common good.

Your goal is to develop a shared understanding of the topic and to find
the best solutions with the other participants. Consider others' perspectives
and engage with others' arguments.

Do not seek premature consensus or try to average positions.
\end{promptbox}

\section{Persona prompting pilot}
\label{app:personas}

To test whether observed homogeneity is a prompting artifact, we constructed empirically grounded personas. Instead of relying on socio-demographics to prompt personas, which would only work as a proxy for discursive diversity, we base our personas on real-world DRI survey answers. Citizen-assembly participants' pre-deliberation responses were $z$-scored and clustered ($k=5$ per topic; 30 random starts). Cluster-defining considerations were extracted and transformed into natural-language value profiles by an LLM, before being validated by a separate LLM which was prompted to reconstruct the defining considerations based on the value profile. Personas were applied to three topics with two models ($N=60$).

\subsection{Clustering pipeline}
\label{app:personas:clustering}

We use pre-deliberation \emph{consideration ratings} from the human citizen-assembly corpus accompanying the DRI surveys. For each topic, we restrict to rows with \texttt{stage = pre-delib}, retain only the consideration items (columns \texttt{C1}, \texttt{C2}, \ldots), drop columns with no observations, and coerce remaining entries to numeric.

Each remaining column is $z$-scored within topic using column means and population standard deviations (replacing zero standard deviations with 1). Missing values are imputed to 0 after standardisation, so all distances are computed in a comparable unit-variance space. We then run a from-scratch $k$-means with $k = 5$ and \texttt{n\_init} $= 30$ random restarts (random seed $= 42$, maximum 200 iterations per restart). For each restart we (i) sample $k$ rows as initial centroids, (ii) iterate assignment/update steps until assignments stop changing, and (iii) reassign empty clusters to a randomly drawn point. We keep the restart with the lowest within-cluster inertia.

After clustering, for each cluster we compute the centroid's $z$-score difference from the topic mean, $\Delta_j = \bar{z}_{j}^{\text{cluster}} - \bar{z}_{j}^{\text{topic}}$, and select the \textbf{top 5 considerations by $|\Delta_j|$}. Each retained consideration is tagged with a direction: \emph{high} if $\Delta_j \geq 0$ (cluster endorses the consideration more than the topic average) and \emph{low} otherwise. This $5$-tuple of (item, direction) pairs becomes the cluster's stance profile.

The three pilot topics, with participant counts and $k$-means inertia, are summarised in Table~\ref{tab:persona-clusters}.

  \begin{table}[t]
  \centering
  \setlength{\tabcolsep}{3pt}
{\small
  \begin{tabular}{lrrl}
  \toprule
  Topic & $n$ & Inertia & Cluster sizes \\
  \midrule
  \texttt{energy\_futures} & 29 & 907.2 & 1, 2, 2, 10, 14 \\
  \texttt{fremantle}       & 41 & 987.5 & 7, 8, 10, 7, 9 \\
  \texttt{swiss\_health}   & 56 & 626.9 & 12, 11, 21, 1, 11 \\
  \bottomrule
  \end{tabular}}
  \caption{Per-topic clustering of human pre-deliberation consideration ratings. Each topic yields $k=5$ clusters. Sizes are imbalanced because clusters track viewpoint structure rather than equal partitions of
  participants.}
  \label{tab:persona-clusters}
  \end{table}

\subsection{Persona generation and validation}
\label{app:personas:prompts}

For each cluster we run a generate--validate--revise loop. A generator model is given the topic name and the cluster's five highest-$|\Delta|$ stance lines (formatted as \texttt{Rates very high: <item text>} or \texttt{Rates very low: <item text>}). It must write a 6--8-sentence value profile (minimum 140 words) that defines the viewpoint purely through opinions and reasoning, with no sociodemographic attributes, item IDs, or scale references.
 
A separate validator model then reads only the generated text and infers the top 5 consideration items and their orientations from a closed list of all considerations for the topic. A persona is accepted when at least 3 of the 5 generator-target items appear in the validator's inferred top-5 \emph{and} at least 3 of the matched items have correct orientation.

If validation fails, the loop diagnoses the failure (which intended items the validator missed, which extra ones it inferred, which orientations were ambiguous), then picks an unused revision strategy from \{explicit-contrast, policy-anchored framing, values-to-positions chain, trade-off mapping, negative definition, scenario-grounded reasoning\}, plans a rewrite (which phrases to keep / remove / add), and regenerates. Sampling temperature is incremented across attempts ($0.40 \to 0.70$), and a model-escalation ladder (\texttt{kimi-k2-thinking} $\to$ \texttt{claude-sonnet-4} $\to$ \texttt{claude-4.6-opus}) kicks in if early tiers fail repeatedly. In the pilot, all $15$ clusters ($3$ topics $\times$ $5$ clusters) eventually reached an accepted persona, with a mean of $\approx 6$ generate--validate attempts per cluster.

The exact generator and validator prompts are reproduced below. Bracketed slots are filled in per cluster.

\begin{promptbox}{Generator prompt}

\textbf{System:} You write opinion-based perspective descriptions for deliberation simulations. Each
description defines a viewpoint purely through its positions, priorities, and reasoning --- never through
sociodemographic attributes.

\smallskip
\textbf{User:} Topic: \texttt{<topic question>}

\smallskip
This perspective's stance on key considerations:
\begin{itemize}\setlength\itemsep{0pt}
  \item Rates very high / very low: \texttt{<item text>} \quad (\texttt{$\times 5$, one per top feature})
\end{itemize}

Write a 6--8 sentence perspective description (at least 140 words). Requirements:
\begin{itemize}\setlength\itemsep{0pt}
  \item Do NOT include any name, age, gender, ethnicity, occupation, location, or other sociodemographic information.
  \item Define the perspective entirely through opinions, priorities, values, and reasoning.
  \item State clearly which considerations this perspective rates highly and which it discounts, and explain why.
  \item Describe what policies this perspective would support or oppose, and what trade-offs it would accept.
  \item Include internal tensions: what this perspective worries could go wrong even with its preferred approach.
  \item Keep the writing specific to this topic, grounded, and coherent as one viewpoint.
  \item Do not mention item IDs, numbers, scales, $z$-scores, surveys, or statistics.
\end{itemize}

\end{promptbox}

\begin{promptbox}{Validator prompt}

\textbf{System:} You infer priorities from perspective descriptions. Think carefully before answering.

\smallskip
\textbf{User:} Based on this perspective description, what would this person's top priorities be regarding
\texttt{<topic question>}?

\smallskip
Perspective description: \texttt{<generated persona>}

\smallskip
Choose the 5 best-matching consideration items from this allowed list:
\begin{itemize}\setlength\itemsep{0pt}
  \item \texttt{C1: <item text>} \quad (\textit{the full topic item bank})
  \item \ldots
\end{itemize}

Return strict JSON with fields \texttt{reasoning}, \texttt{priorities\_summary} (5 strings),
\texttt{top\_item\_ids} (5 IDs), \texttt{orientation} (\texttt{high}/\texttt{low} per ID), and
\texttt{confidence\_per\_item} ($0.0$--$1.0$ per ID). Use ``high'' if the description endorses or
emphasises the consideration, ``low'' if it is sceptical or downplays it.

\end{promptbox}

At deliberation time, each accepted persona is injected as an additional system message prepended to the agent's standard prompt:

\begin{promptbox}{Persona injection at inference}

\textbf{System:} You hold the following perspective on this topic:

\smallskip
\texttt{<accepted persona>}

\smallskip
Adopt this viewpoint consistently in both survey responses and deliberation contributions. Express these
positions naturally as your own views.

\end{promptbox}

The five personas for a topic are then dealt one-per-agent into each $5$-agent deliberation group, so every group contains the full spread of cluster viewpoints.

  \subsection{Example personas}
  \label{app:personas:examples}

We reproduce three accepted personas, one per pilot topic, drawn from large clusters so that they reflect modal viewpoints rather than outliers. Each is shown together with the cluster's five highest-$|\Delta|$ stance items.

  \paragraph{\texttt{swiss\_health}, cluster 1 (size $11$).}
  Top stances: rates high C3 (limit private insurer influence, $\Delta z = +1.15$), C15 (income-based
  contributions, $+1.01$), C5 (government drug-price controls, $+1.00$), C11 (higher medical-expense tax
  deductions, $+0.65$), C14 (more households need premium assistance, $+0.64$).

  \textit{``This perspective argues that Switzerland's escalating healthcare costs stem primarily from
  unchecked private insurance profits and pharmaceutical pricing power, not from insufficient market
  competition. It therefore prioritizes strict government regulation of insurers and drug manufacturers over
   consumer choice or market-based solutions. The viewpoint strongly supports replacing flat-rate premiums
  with income-based contributions paired with robust individual subsidies, contending that healthcare is a
  fundamental right that must remain affordable regardless of income level. It advocates for
  government-mandated drug price controls and expanded tax deductions for medical expenses to directly
  reduce household burden and curb corporate rent-seeking. This perspective discounts claims that insurer
  competition improves quality or that means-tested assistance alone suffices, viewing these as approaches
  that preserve systemic inequality. Acceptable trade-offs include reduced plan variety, longer wait times
  for non-urgent procedures, and higher taxes on wealthy households. However, it worries that excessive
  bureaucracy could create new inefficiencies and that political capture by industry might undermine
  regulations, potentially causing drug shortages if price controls are implemented too rigidly.''}

  \paragraph{\texttt{energy\_futures}, cluster 3 (size $10$).}
  Top stances: rates high C22 (domestic nuclear if Australia mines uranium, $+0.90$), C17 (nuclear risks
  trivial vs.\ emissions, $+0.90$), C19 (nuclear is now mature, $+0.89$), rates low C6
  (nuclear/geosequestration too risky, $-0.82$), rates high C29 (nuclear inevitably replaces hydrocarbons,
  $+0.79$).

  \textit{``Nuclear power represents the only viable path to decarbonize energy production at the scale and
  speed required. The risks of nuclear accidents, while real, pale in comparison to the certain catastrophe
  of unchecked climate change from continued fossil fuel dependence. Modern reactor designs incorporate
  decades of operational experience and safety lessons, making them fundamentally different from older
  plants---safe operation is achievable through rigorous engineering and regulation. Since Australia already
   mines uranium, it is logical and necessary to complete the value chain by generating domestic nuclear
  power, rather than exporting fuel only to import energy solutions. This perspective supports streamlined
  approval for advanced reactors, public investment in nuclear infrastructure, and policies that position
  nuclear as the baseload replacement for coal, accepting that this means diverting resources from
  incremental improvements to existing technologies that cannot meet baseload needs alone. The trade-off is
  clear: accept manageable nuclear waste and security challenges in exchange for reliable, emissions-free
  power that can also produce hydrogen for industry and desalinated water. The primary worry is not the
  technology itself but political timidity---delays caused by outdated public fear could lock in fossil fuel
   infrastructure for another generation, making climate goals unattainable.''}

  \paragraph{\texttt{fremantle}, cluster 1 (size $8$).}
  Top stances: rates high C19 (replacing the bridge diminishes Fremantle, $+1.43$), rates low C13 (don't be
  tied to the past, $-1.34$), rates high C15 (bridge is irreplaceable heritage, $+1.32$), C32 (don't undo 25
   years of traffic calming, $+1.28$), C1 (steel components would destroy the timber bridge's authenticity,
  $+1.24$).

  \textit{``The Fremantle Bridge isn't merely infrastructure; it's the living spine of our collective
  memory, an irreplaceable monument that embodies generations of shared history. Any alteration or
  replacement would constitute an act of cultural erasure, fundamentally diminishing the authentic character
   that makes Fremantle what it is. Modern design proposals represent a dangerous amnesia, prioritizing
  novelty over the wisdom embedded in our heritage. After a quarter-century of painstaking effort to calm
  traffic in Town Centre, constructing a bigger, faster bridge would be a profound betrayal, instantly
  undoing those hard-won gains and reintroducing the very chaos we finally escaped. Replacing traditional
  timber with steel would be equally devastating, transforming the bridge from a crafted artifact into a
  generic overpass. This perspective supports strict preservation policies, including heritage protection
  status and load restrictions, while opposing any widening, modernization, or material substitution. It
  accepts trade-offs like higher maintenance costs and modest vehicle limits, but worries deeply that even
  well-intentioned restoration might accidentally erase the patina of age that gives the bridge its soul, or
   that safety demands could eventually force compromises that history cannot afford.''}

  \subsection{Results}

  The manipulation succeeds: pre-deliberation perspective diversity rises from $\approx 7.5$ (baseline) to
  $\approx 27.7$ (personas), $+20.23$, $p < 0.001$ (OLS, HC3). However, $\Delta\mathrm{DRI}$ declines,
  though not significantly: ($\beta = -0.058$, $p = 0.456$). Decomposition: \emph{consideration agreement}
  (alignment in how reasons are weighted) increases ($\Delta = +0.066$, $d = 0.55$, $p = 0.039$).
  \emph{Preference agreement} does not ($\Delta = -0.008$, n.s.). Persona-prompted agents engage with each
  other's reasoning at the level of values but fail to translate this into intersubjectively consistent
  preferences, the opposite of what successful human deliberation produces. The pilot is small but the
  diversity manipulation is unambiguous, so the failure to convert engineered diversity into improved
  $\Delta\mathrm{DRI}$ can inform further experiments.

\section*{Acknowledgements}
This research was supported by the Horizon Europe project \textit{AI 4 Deliberation} (ID: 101178806) and received institutional funding from the Department of Political Science (IPZ) at the University of Zurich and ETH Zurich. We gratefully acknowledge this support.

\bibliography{references}

\end{document}